\documentclass[12pt]{article}
\usepackage{graphicx}

\begin{document}

\begin{center}
\Large {\bf `Thermodynamics' of Minimal Surfaces \\ and Entropic Origin of Gravity}
\end{center}

\bigskip
\bigskip

\begin{center}
D.V. Fursaev
\end{center}

\bigskip
\bigskip

\begin{center}
{\it Dubna International University \\
     Universiteskaya str. 19\\
     141 980, Dubna, Moscow Region, Russia\\

  and\\

  the Bogoliubov Laboratory of Theoretical Physics\\
  Joint Institute for Nuclear Research\\
  Dubna, Russia\\}
 \medskip
\end{center}

\bigskip
\bigskip

\begin{abstract}
Deformations of minimal surfaces lying in constant time slices in static space-times are studied.
An exact and universal formula for a change of the area of a minimal surface under shifts of
nearby point-like particles is found. It allows one to introduce a local temperature
on the surface and represent variations of its area in a thermodynamical form by assuming that
the entropy in the Planck units equals the quarter of the area.
These results provide a strong support to a recent hypothesis that gravity has an entropic
origin, the minimal surfaces being a sort of holographic screens.
The gravitational entropy also acquires a definite physical meaning related to quantum entanglement of fundamental  degrees of freedom across the screen.
\end{abstract}

\newpage

\section{Introduction}

The fact that gravity is an emergent phenomenon dates back to ideas of the last
century \cite{Sakharov:1967pk}. A renewal of the interest
to this point of view in last years has been motivated by attempts to find a statistical explanation of
the Bekenstein-Hawking entropy, see e.g. \cite{Jacobson:1994iw,Fursaev:2004qz,Frolov:1998vs} and
references therein. A possible source of the entropy are quantum correlations of underlying microscopical degrees of freedom across the black hole horizon \cite{Sr:93,BKLS,FrNo:93}.

By taking the black hole case as a guide
a number of arguments have been presented in \cite{Fursaev:2007sg} that the entanglement entropy of
fundamental degrees of freedom lying in a constant time slice and
spatially separated by a surface $\cal B$
is
\begin{equation}\label{1}
S({\cal B})={{\cal A}({\cal B}) \over 4G}~~.
\end{equation}
Here $G$ is the Newton coupling and
$\cal A$ is the area of ${\cal B}$. Thus, (\ref{1}) has the Bekenstein-Hawking form.
Equation (\ref{1}) holds in the semiclassical approximation if the low-energy limit of the
fundamental theory is the Einstein gravity.

For realistic condensed matter systems the entanglement entropy associated to spatial separation of the system is a non-trivial function of microscopical parameters.
Its calculation is technically involved and model dependent. The remarkable consequence of (\ref{1})
is that the entanglement entropy in quantum gravity may not depend on a microscopical
content of the theory, it is determined solely in terms of geometrical characteristics of the surface
and low-energy gravity couplings.

Another feature established in \cite{Fursaev:2007sg} is related to the shape of the
separating surface. Because $S({\cal B})$ includes contributions of all fundamental degrees of freedom quantum
fluctuations of the geometry should be taken into account in a consistent way.
For static space-times this requires that $\cal B$
is minimal surface, i.e. a surface with a least area.
A relevant physical example of a minimal surface is the intersection of a constant time slice and
the horizon of a stationary black hole. Thus, the Bekenstein-Hawking entropy
can be considered as a particular case of the entanglement entropy (\ref{1}).

The fact that $S({\cal B})$ is a macroscopic quantity which obeys certain dynamical laws
points to similarity with a thermodynamical entropy. A natural question is whether the entanglement
on the fundamental level admits a form of thermodynamical laws.

A first step in this direction was done in \cite{Fursaev:2007sg}. A calculation made there in the
weak field approximation shows  that a shift by a distance $l$ of a test particle with a mass $m$ out of the
minimal surface  results in the entropy change
\begin{equation}\label{0}
\delta S({\cal B})=-\pi m~l~~.
\end{equation}
A work needed to drag the
particle by the background gravitational field is also proportional to $l$.
Now by following the recent observation by Verlinde \cite{Verlinde:2010hp} one can relate the entropy change (\ref{0}) and the work,
the relation being an analog of the first law of thermodynamics.
As we show this yields a local temperature on the surface. The temperature is proportional to the product of
the acceleration of a static observer near the surface and the normal vector to $\cal B$.

A remarkable hypothesis of \cite{Verlinde:2010hp} is that
gradients of the entropy of fundamental
microscopical degrees of freedom in an underlying quantum gravity theory determine gradients of the gravitational field. To cut it short the force of gravity is an entropic force.
The hypothesis is based on a number of assumptions for so called `holographic screens' which
store an information about fundamental microstates ('bits') in such a way that a related entropy
is proportional to the area of the screen. A variational formula for the entropy
of the screen
under the action of a point-like particle is postulated in the form like (\ref{0}) and plays a key role
in the arguments.

The main goal of this paper is to show that (\ref{0}) follows from the properties of minimal surfaces
in general static space-times and to establish an analogy between dynamics of minimal surfaces
and thermodynamical systems. Minimal surfaces, therefore, may be considered as a sort of
holographic screens with a well-understood dynamics. The form of this dynamics yields a
strong support to the hypothesis of \cite{Verlinde:2010hp}.
A possible relation of an entropic force and quantum entanglement has been  discussed recently by a number of authors, see e.g. \cite{Myung:2010rz,Lee:2010fg}, but not in the context of minimal surfaces.

In Section 2 we consider a weak field limit of gravity theory and study dynamics of minimal surfaces
caused by shifts of test point-like particles located close to the surface. To make
a connection with the ideas of \cite{Verlinde:2010hp} we start with a system
which consists of two infinite non-intersecting planes and a massive source in between.
The planes play the role of two components of a holographic screen.
The Komar integral on the each plane equals half
of the mass of the source. From this example we move to discussion of properties
of a single minimal surface with the topology of a plane, introduce a local temperature on the surface
and an analog of the first law.

The aim of next sections  is to extend these results beyond the weak field approximation. In Section 3 we prove (\ref{0})
for the entropy of a minimal surface in a slice in a static space-time which is a solution
to the Einstein equations in a vacuum. The key property used here is an approximate isometry of the slice
in the direction orthogonal to the surface.

Thermodynamical interpretation of the behavior of minimal surfaces in static space-times
is discussed in Section 4. An advantage of taking holographic screens as minimal surfaces is in the clear physical meaning of the corresponding entropy. This opens a possibility to
study further questions which we briefly describe.

We use units where $\hbar=c=k_B=1$.

\section{Dynamics in a weak field approximation}

\begin{figure}[h]
\begin{center}
\includegraphics[height=5cm,width=4.5cm]{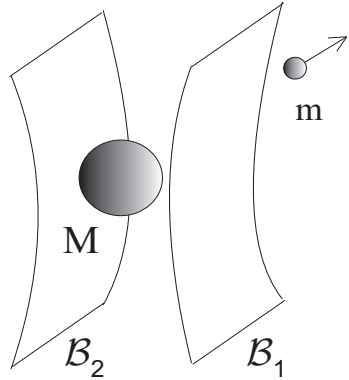}
%\v_space{0.4cm}
\caption{\small{The figure shows two minimal surfaces, ${\cal B}_1$, ${\cal B}_2$,
around a massive source with the mass $M$
and a test particle with the mass $m$. Shifts of the test particle result in a
finite change of the area of the minimal surfaces.}}
\label{f1}
\end{center}
\end{figure}

Consider a gravitating source with the mass $M$.
The  geometry around the source is a static four-dimensional space-time $\cal M$.
Constant-time slices in $\cal M$ are denoted by $\Sigma$.
We need a holographic screen  around the source which is a minimal surface in $\Sigma$. A closed minimal surface
exists only around a black hole and it coincides with the black hole horizon.
Therefore, we consider a
screen which consists of two non-intersecting infinite components, ${\cal B}_1$, ${\cal B}_2$, with the
source located in between. We assume that ${\cal B}_k$ are minimal surfaces  in $\Sigma$
which have topology of a plane, see Fig. \ref{f1}.
The orientation of the surfaces can be specified by
conditions at asymptotic infinity.

It is convenient to start with the weak field approximation when
the space-time
metric is
\begin{equation}\label{2}
ds^2=-\left(1+2\phi\right)dt^2+\left(1-2\phi\right)(dx^2+dy^2+dz^2)~~~.
\end{equation}
Here  $\phi(r)=-{MG \over r}$, $r=\sqrt{x^2+y^2+z^2}$, is the gravitational potential of the source
located at $r=0$.
In the weak field approximation ${\cal B}_k$ are just parallel planes.
One can direct the $z$ axis orthogonally to the surfaces and fix
coordinates of the surfaces $z_k$. It is assumed that $|z_k|\gg MG$.

For the each surface one can define the following integral:
\begin{equation}\label{4a}
{\cal E}({\cal B}_k)={1 \over 4\pi G} \int_{{\cal B}_k}d\sigma w_n~~~,
\end{equation}
where $d\sigma=dxdy$ and $w_n=\partial_n\phi$ is a normal component of the acceleration of a static observer located at the surface. The normal vectors to ${\cal B}_k$ are
directed opposite to the source. The sum of the two integrals
is the Komar mass which coincides with the mass of the source,
\begin{equation}\label{4b}
{\cal E}={\cal E}({\cal B}_1)+{\cal E}({\cal B}_2)=M~~
\end{equation}
because each ${\cal E}({\cal B}_k)$ equals $M/2$.
From this point of view the two-component screen is equivalent to a closed screen around the source.

Consider now the entropy (\ref{1}) related to the surfaces.
It is not difficult to see that the area of the surface in the weak field approximation
is given by the integral
\begin{equation}\label{3}
{\cal A}({\cal B}_k)=\int dxdy\left(1-2\phi(r_k)\right)~~~,
\end{equation}
where $r_k=\sqrt{x^2+y^2+z_k^2}$. To avoid divergences in (\ref{3}) one may assume that the surfaces have a large but finite size.
Then according to (\ref{1})
\begin{equation}\label{4}
S_k={{\cal A}({\cal B}_k) \over 4G}=\bar{S}_k+\int d\textit{s}(r_k)~~~,
\end{equation}
where $\bar{S}_k$ is a constant, an entropy associated to a plane, and
\begin{equation}\label{5}
d\textit{s}(r)=-{1 \over 2G}\phi(r)d\sigma~~.
\end{equation}
By following \cite{Verlinde:2010hp} we define the number $dN$ of degrees of freedom on an element
of the screen with the area $d\sigma$ as
\begin{equation}\label{6}
d \textit{s}(r)=-{\phi(r) \over 2}dN~~
\end{equation}
and come to equation
\begin{equation}\label{7}
dN={d\sigma \over G}~~.
\end{equation}
This coincides with the result by \cite{Verlinde:2010hp} but in our case (\ref{7}) follows immediately from the
the Bekenstein-Hawking relation (\ref{1}). A relevance of (\ref{1}) for the entropy
on a holographic screen has been also pointed out in \cite{Kiselev:2010mz}.

In this almost trivial setting one can study dynamics of minimal surfaces
under the action of a test point-like particle.  The gravitational field of the particle changes
the shape of the surfaces. The area of the surface
becomes
\begin{equation}\label{8}
{\cal A}({\cal B}_k)=\int d\sigma \left[1-2\phi(r_k)-2\phi'(r'_k)\right]~~~.
\end{equation}
Here $\phi'(r'_k)=-m G / r'_k$, $r'_k=\sqrt{(x-x_0)^2+(y-y_0)^2+(z_0-z_k)^2}$, is the gravitational potential
of the particle on the surface, $m$ is the mass of the particle, and $x_0,y_0,z_0$ are its
coordinates ($z_0>z_k$). A variation of the area of the surface under shifts of the particle is finite \cite{Fursaev:2007sg},
\begin{equation}\label{9}
\delta {\cal A}({\cal B}_k)=\delta z_0\int dxdy{2mG  (z_0-z_k) \over ((x-x_0)^2+(y-y_0)^2+(z_0-z_k)^2)^{3/2}}=
-4\pi mG l~~~,
\end{equation}
where $l=\delta z_0$ is the change of the distance between the particle and the surface.
Displacements $\delta x_0$, $\delta y_0$ of the particle along the planes do not change the area because of
a translational invariance.
According to (\ref{19}) the area of the surface increases if the particle moves towards the surface
$(\delta z_0 < 0)$.
The closer the particle to the surface the larger distortion of its shape.

If one restores all dimensional
constants the variational formula for
the entanglement entropy of the whole screen takes the form
\begin{equation}\label{10}
\delta S=\delta S_1+\delta S_2=-2\pi {mc \over \hbar} l~~~.
\end{equation}
It is assumed here that the particle moves out of the screen.
In \cite{Verlinde:2010hp} the variational formula (\ref{10}) has been postulated for surfaces of equal potentials.

Note that the situation is different when the particle
moves between the surfaces. Then it is a part of the system and the Komar
energy (\ref{4b}) is increased
by the mass of the particle. Because
the particle moves from one surface to the other variations $\delta S_k$ have opposite signs.
Hence, when
the particle moves staying inside the screen the whole entropy does not change, $\delta S=0$.  There is an obvious analogy here between the two-component screen and a
black hole horizon. The analogy can be strengthen if we recall that (\ref{1}), as
the entropy of entanglement, appears when the information about the states located inside the
screen is not accessible for an outside observer.

We now return to the case when the particle is outside the screen.
The gravitational field of a test particle moving in a vacuum changes a distribution of fundamental
degrees of freedom and, therefore, affects the way they are entangled.
On thermodynamical grounds a work $\delta W(x)$ which is required to change the state of
underlying degrees of freedom on the holographic screen by dragging the test particle (located at a point with the coordinates
$x$) is
\begin{equation}\label{11}
\delta W(x)=-T(x)\delta S~~~,
\end{equation}
where $\delta S$ is the variation of the entropy (\ref{10}) and $T(x)$ is a multiplier. One can then
define an entropic force as force which performs the work (\ref{11})
\begin{equation}\label{12}
\delta W(x)=-F(x)l~~~.
\end{equation}
Here $F=-m w_n(x)$ is an external force and the work (\ref{12}) is positive if the particle moves
in the direction opposite to the gradient of the gravitational field. Expression for the multiplier follows from (\ref{10}), (\ref{11}),
\begin{equation}\label{13}
T(x)={w_n(x) \over 2\pi}~~~.
\end{equation}
By using analogy of (\ref{11}) and the first
law of thermodynamics $T(x)$ can be associated to a temperature, its expression being exactly the same
as the result of \cite{Verlinde:2010hp}. Note that entropic force balances the component of force of gravity
normal to the screen. Thus, in the Newtonian theory one may say that the gravity force has an
entropic nature. The entropy increases in the direction of gradients of the gravitational potential.

For the following we need a definition of the temperature on the each surface.
The components of the screen are equivalent, each carries the energy $M/2$, see (\ref{4a}), and
acquires the same entropy variation, $\delta S_k=-\pi m l$.
Therefore, the work required to change the state
of the single component is just a half of the total work,
\begin{equation}\label{14}
T_k(x)\delta S_k=-\frac 12 \delta W(x)~~.
\end{equation}
A local temperature on ${\cal B}_k$ can be defined
in the limit when the test particle approaches the surface
\begin{equation}\label{15}
T_k(x)=\left. {w_n(x) \over 2\pi} \right|_{x\in {\cal B}_k}~~~.
\end{equation}
In the weak field approximation $T$ and $T_k$ coincide if the distance between the surfaces is much smaller than the distance of the test particle to the source.

Having this definition and the definition of the number of degrees of freedom on an element
of the screen (\ref{9}) we can rewrite the Komar integral (\ref{4a}) for a single surface as
\begin{equation}\label{16}
{\cal E}({\cal B}_k)={1 \over 2} \int_{{\cal B}_k} T_k dN~~.
\end{equation}
The relation above looks as an `equipartition' formula for an energy (see discussion in  \cite{Padmanabhan:2009kr,Verlinde:2010hp}). This gives another justification of choosing
a thermodynamic law for the single surface in the form (\ref{14}). More arguments in favor of
(\ref{14}) follow in Section 4.

\section{Dynamics in general background fields}

Let us show that a minimal surface in a general static space-times allows the same variational formula
(\ref{10}) under the action of the gravitational field of a test particle.

We consider a static asymptotically flat space-time $\cal M$  with a metric $g_{\mu\nu}$
which is a solution to the Einstein
equations.
The dimensionality $n$ of $\cal M$ is not fixed.
The space-time may describe either geometry around a black hole or a matter
source located in a compact region.  We are interested in constant time slices $\Sigma$ on $\cal M$
which are orthogonal to a time-like Killing vector field $\xi^\mu$.
The metric on $\cal M$ can be written in the form
\begin{equation}\label{19}
ds^2=-Bdt^2+dl^2=-B(x)dt^2+g_{ab}(x)dx^adx^b~~,
\end{equation}
where components do not depend on time, $a,b$ are spatial indexes, and $dl^2$ is the metric on a slice $\Sigma$.

Let $\cal B$ be a minimal least area surface in $\Sigma$.
The position of $\cal B$ is determined by a set of equations $X^a=X^a(y)$, where $y^i$ are coordinates on $\cal B$. The metric induced on $\cal B$ is $\gamma_{ij}=g_{ab}X^a_{,i}X^b_{,j}$.
The position of $\cal B$ can be fixed by conditions
at asymptotic infinity. If one places a test-particle near the surface the metric on $\cal M$ becomes
$g_{\mu\nu}+h_{\mu\nu}$ and the area of the surface acquires a variation which we denote as
${\cal A}'({\cal B})$. A trivial calculation shows that
\begin{equation}\label{23}
{\cal A}'({\cal B})=\frac 12 \int_{\cal B}\sqrt{\gamma} d^{n-2}y~ \gamma^{ij} ~X^a_{,i}X^b_{,j}h_{ab}~~.
\end{equation}
The perturbations $h_{\mu\nu}$ can be considered
in the linear order because the mass of the test particle is assumed to be small. The surface may also change its position
but this causes no effect on ${\cal A}'$ because the surface is minimal.

We allow a cosmological constant $\Lambda$ in the Einstein equations but assume that $\cal B$
does not intersect location of a gravitating source.
By following the standard procedure one gets from the Einstein
equations the equations for perturbations
\begin{equation}\label{17}
L\bar{h}_\mu^\nu=16\pi G~t_\mu^\nu~~,
\end{equation}
$$
L\bar{h}_\mu^\nu=-\nabla^2~\bar{h}_\mu^\nu-2R_{\mu\rho}~^{\nu\lambda}~\bar{h}^\rho_\lambda-{8\Lambda \over n-2}\bar{h}_\mu^\nu~~,
$$
where $R_{\mu\rho\nu\lambda}$ is the Riemann tensor of $\cal M$, and
\begin{equation}\label{18}
\bar{h}_\mu^\nu=h_\mu^\nu-\frac 12 h \delta_\mu^\nu~~,
\end{equation}
$h=h_\mu^\mu$. As usually, the gauge freedom has been used to set the condition $\nabla^\lambda \bar{h}_{\lambda\nu}=0$.
The stress-energy tensor of the test particle which is at rest in coordinates (\ref{19}) is
\begin{equation}\label{20a}
t^\nu_\mu=m \delta^{(n-1)}(x,x_0)u^\mu u_\nu~~~,
\end{equation}
where $u^\mu=\xi^\mu/\sqrt{B}$ is the velocity and $m$ is the mass of the particle.
The energy density  measured by an observer with the velocity vector $u^\mu$ is $t_{\mu\nu}u^\mu u^\nu$.
The energy density for static observers is $m \delta^{(n-1)}(x,x_0)$.
We choose the normalization
$$
\delta^{(n-1)}(x,x')=\delta^{(n-1)}(x-x')/\sqrt{\det g_{ab}(x)}~~,
$$
\begin{equation}\label{20b}
\int_{\Sigma} t_{\mu\nu}u^\mu u^\nu \sqrt{\det g_{ab}}~ d^{n-1}x=m~~~
\end{equation}
to ensure that integration of this density over the slice $\Sigma$ yields the mass of the particle.

In (\ref{17}) the wave operators on static solutions ($\partial_t \bar{h}_{\mu\nu}=0$) are
reduced to covariant Laplacians on $\Sigma$ and a number of terms
which depend on the acceleration $w^\mu=\nabla_u u^\mu$.
For instance,
\begin{equation}\label{21}
\nabla^2~ \bar{h}_0^0=\Delta ~\bar{h}_0^0-w^a\partial_a \bar{h}_0^0+2w^2~ \bar{h}_0^0-2w_a w^b ~\bar{h}_b^a~~,
\end{equation}
where $\Delta$ is a scalar Laplacian on $\Sigma$.

We consider the perturbations in a domain of the minimal surface $\cal B$ which is located in a vicinity of the particle.
One expects that in this region the operator $L$ in (\ref{17}) can be approximated by
the covariant Laplacians $-\Delta$. This happens because spatial changes of the perturbations are
so large that the curvature, $\Lambda$, and acceleration terms in $L$ can be neglected.
In the coordinates (\ref{19}) the only non-zero component of stress tensor (\ref{20a}) is $t^0_0$ and
in the given approximation (\ref{17}) yield the only non-vanishing perturbation
\begin{equation}\label{22c}
\bar{h}_0^0(x)=16\pi G m~D(x,x_0)~~,
\end{equation}
\begin{equation}\label{22d}
\Delta_x D(x,x')=\delta^{(n-1)}(x,x')~~.
\end{equation}
This single perturbation, if one uses (\ref{18}) and (\ref{23}), results
in the following variation of the area
\begin{equation}\label{23-a}
{\cal A}'({\cal B})=
-
\frac 12 \int_{\cal B}\sqrt{\gamma} d^{n-2}y~ \bar{h}^0_0~~.
\end{equation}

The approximation above does not mean that the slice $\Sigma$ is considered as a flat
space and the minimal surface as a plane. Let us show that despite the fact that $\Delta$ is a
curved Laplacian the dynamic of the minimal surface under the action
of the particle is exactly the same as in the weak field approximation.

For further discussion it is convenient to use on $\Sigma$ the
Gaussian normal coordinates which start on $\cal B$. The metric can be written as
\begin{equation}\label{24}
dl^2=dz^2+\gamma_{ij}(z,y)dy^idy^j~~.
\end{equation}
The position of $\cal B$ is $z=0$ and $\gamma_{ij}(0,y)$
coincides with the metric on $\cal B$. The
extrinsic curvature tensor for $\cal B$ in these coordinates is
$k_{ij}(y)=-\frac 12\partial_z\gamma_{ij}(z,y)$. Because the surface is a minimal the trace
$k_{ij}\gamma^{ij}(y)$ is vanishing, and the positivity of $\gamma^{ij}$ implies that
\begin{equation}\label{25}
\partial_z\gamma_{ij}(0,y)=0~~.
\end{equation}
It is important that (\ref{25}) holds globally along the surface, i.e. there is
a translational invariance along the $z$ direction in a narrow layer near $\cal B$.
In this layer one can write
\begin{equation}\label{27}
D(x,x')=D(z-z';y,y')~~,
\end{equation}
\begin{equation}\label{36}
D(z;y,y')={1 \over 2\pi}\int_{-\infty}^\infty dk e^{ikz} \tilde{D}_k(y,y')~~,
\end{equation}
where $\tilde{D}_k(y,y')$ is an operator on $\cal B$ which is a solution to
\begin{equation}\label{37}
(\tilde{\Delta}-k^2)\tilde{D}_k(y,y')=\delta^{(n-2)}(y,y')~~,
\end{equation}
and $\tilde{\Delta}$ is a scalar Laplacian on $\cal B$.

Let $(z_0,y_0^i)$ be coordinates of the test particle on $\Sigma$, $z_0>0$.
If the particle moves to a nearby
point with coordinates $(z_0+l,y_0^i+l^i)$
the area of the minimal surface changes as
\begin{equation}\label{29}
\delta {\cal A}({\cal B})=\delta {\cal A}'({\cal B})=\delta_{\parallel} {\cal A}'({\cal B})
+\delta_{\perp} {\cal A}'({\cal B})~~,
\end{equation}
\begin{equation}\label{29a}
\delta_{\parallel} {\cal A}'({\cal B})=-
\kappa ~l^i
\int_{\cal B}\sqrt{\gamma} d^{n-2}y~ {\partial \over \partial y_0^i} D(z-z_0;y,y_0)~~,
\end{equation}
\begin{equation}\label{29b}
\delta_{\perp} {\cal A}'({\cal B})=-
\kappa~l
\int_{\cal B}\sqrt{\gamma} d^{n-2}y~ {\partial \over \partial z_0} D(z-z_0;y,y_0)~~,
\end{equation}
where $\kappa=8\pi G m$. Here we have used (\ref{22c}) and (\ref{23-a}) and considered a linear order
in coordinate shifts.

The variation is composed
of two parts, $\delta_{\parallel} {\cal A}'({\cal B})$, $\delta_{\perp} {\cal A}'({\cal B})$, which
correspond to shifts in directions parallel and normal to the surface. Let us consider first parallel
variations. As in the previous section, we introduce an infrared regularization by assuming that
$\cal B$ has a large but finite size. Then $\tilde{\Delta}$ on $\cal B$ has a
discreet  spectrum specified by some boundary conditions, the Drichlet or Neuman conditions, for example. Let
$\varphi_\lambda(y)$ be a complete set of eigenmodes
$\tilde{\Delta}\varphi_\lambda=\lambda \varphi_\lambda$, normalized as
\begin{equation}\label{38}
\int_{\cal B}\sqrt{\gamma} d^{n-2}y ~\varphi^*_\lambda(y) \varphi_{\lambda'}(y)=\delta_{\lambda,\lambda'}~~.
\end{equation}
The solution to (\ref{37}) can be written as
\begin{equation}\label{39}
\tilde{D}_k(y,y')=\sum_\lambda{\varphi_{\lambda}(y)\varphi^*_\lambda(y')  \over \lambda-k^2}~~~.
\end{equation}
For the Neuman boundary conditions the operator $\tilde{\Delta}$ has a non-trivial
constant eigenfunction which
is a zero mode.  For
the Dirichlet conditions non-trivial constant modes are absent. Therefore, (\ref{38}) can be used to
write
\begin{equation}\label{40}
\int_{\cal B}\sqrt{\gamma} d^{n-2}y ~\varphi_\lambda(y)=c\delta_{\lambda,0}~~,
\end{equation}
where $c=\sqrt{\mbox{vol}({\cal B})}$ for Neuman conditions and $c=0$ for the Dirichlet conditions. One concludes
that
\begin{equation}\label{41}
\int_{\cal B}\sqrt{\gamma} d^{n-2}y ~\partial_{y'}\tilde{D}_k(y,y')=0~~.
\end{equation}
As a consequence of (\ref{36}), (\ref{41}) parallel
shifts (\ref{29a}) do not change the area of the minimal
surface, $\delta_{\parallel} {\cal A}'({\cal B})=0$.

Let us turn to the normal part (\ref{29b}) of the variation. It can be written as
a surface integral of the normal derivative,
\begin{equation}\label{29c}
\delta_{\perp} {\cal A}'({\cal B})=
\kappa~l
\int_{\cal B}\sqrt{\gamma} d^{n-2}y~ {\partial \over \partial z} D(z-z_0;y,y_0)~~.
\end{equation}
When the infrared regularization is removed (\ref{29c})
can be converted into an integral of $\Delta D$ over $\Sigma$.  After that one can use (\ref{22d})
to get for the area variation
\begin{equation}\label{30}
\delta_{\perp} {\cal A}'({\cal B})=-\frac 12 \kappa~l=-4\pi m G l~~.
\end{equation}
Note that the volume integral is equivalent to two surface terms, one on the surface $\cal B$ and
another on a surface located at some coordinate $z>z_0$. The two surface terms for a point-like source
are equivalent.

We conclude that variation of the area of a minimal
surface is proportional to the change of the distance
between the test particle and the surface, the form of the variation being exactly
the same as in the weak field approximation. The area of the surface increases if the particle moves
to the surface, $l<0$.

\section{Discussion}

As has been suggested in \cite{Verlinde:2010hp} gradients of the entropy of fundamental
microscopical degrees of freedom in an underlying quantum gravity theory might determine the gradients of the gravitational
field. Together with the fact that the gravitational field equations appear to follow from a maximization of the entropy \cite{Padmanabhan:2009vy,Verlinde:2010hp} one gets the strong evidence that gravity  is an emergent phenomenon.

The results above are based on a number of postulates. In particular, one needs to clarify the physical meaning of
the entropy.
The key element in the reasonings of \cite{Verlinde:2010hp} are holographic screens
which store an information, the total number of fundamental bits being proportional
to the area of a screen. A remarkable fact that laws of gravity take the form of
thermodynamical relations requires screens to be dynamical
surfaces which change their shape under the action of the matter.  Verlinde's suggestion was to take the screens as equipotential surfaces. This is a straight way to relate entropy variations to gradients of the
gravitational potential. However, it is not proved that dynamical properties of equipotential surfaces lead to the variational formula (\ref{10}).

One of the avenues in looking for alternative dynamical screens is to study different kinds
of horizons, as black hole mechanics teaches us. Some results \cite{Cai:2010sz} do indicate
that, for example, trapping horizons allow an analog of the entropic force.
The other option is to try to derive dynamical equations by studying lensing effects for
a flow of trajectories passing through the screen. At the present moment, however,
it is not clear which kind of trajectories has to be chosen, see \cite{Piazza} for some suggestions.

In previous sections we considered the screens as co-dimension 2 minimal surfaces in static space-times.
In a quantum gravity theory a minimal surface encodes an information about entanglement of microscopical
degrees of freedom spatially separated by this surface.
The property that the entropy scales as the area of the surface is a physical fact which has been supported by computations in many-body systems \cite{Sr:93,BKLS}.
As for the minimal shape of the surface it is formed under quantum fluctuations of the geometry  \cite{Fursaev:2007sg}.

Not only the definite physical meaning of the entropy but also the possibility to
derive explicit variational formula are among the important features of the minimal surfaces.
Equations (\ref{29}) and (\ref{30}) result in formula (\ref{0}) which yields the change of the entropy (\ref{1})
under the shift of the test particle with respect to the screen by the distance $l$.
Equation (\ref{0}) does not depend on the form of background solution
and dimensionality
of space-time.
It is the same relation as in the weak field approximation and further discussion
of thermodynamical properties
repeats arguments of Section 2.

Let us assume that $\Lambda=0$ and the test particle is suspended on a weightless rigid string near a minimal surface,
the end of the string being in the region where the gravitational effects can be neglected.
The surface is supposed to be between a massive source and the particle. Therefore,
the acceleration vector of the particle $w^\mu$ is directed toward the surface.
The force applied to the end of the string to hold the particle in a fixed position is $-m\sqrt{B}w^\mu$,
see  \cite{FroNo}. A distant observer can pull the end of the string
to change position of the particle in the direction normal to the surface.
The work done in this process is
$\delta W(x)= m \sqrt{B} w_n(x)l$, where $w_n(x)=w^\mu(x)n_\mu(x)$
and $n_\mu$ is the unit vector normal to $\cal B$.
The work is positive, if $l>0$, i.e. if the particle is shifted in the direction opposite to the
acceleration, as it should be for a work done on the system by external forces. One can now
relate the work and the variation of the entropy,
\begin{equation}\label{32}
T(x)\delta S=-\frac 12 \delta W(x)~~~.
\end{equation}
The coefficient $T(x)$ has the meaning of a local temperature
measured by a distant observer. With the help of (\ref{0}) one finds
\begin{equation}\label{34}
T(x)=\sqrt{B(x)}{w_n(x) \over 2\pi}~~~.
\end{equation}
In (\ref{32}) we took into account that the work should be distributed
between two components of the holographic screen. In our case the second component is not considered.
A single component with the topology of a hyperplane corresponds to half of the work because it carries half of the information about the entire system, see Section 2.
This can be seen again from the fact that the `energy' of the surface which is defined by the Komar
integral
\begin{equation}\label{35}
{\cal E}({\cal B})={1 \over 4\pi G} \int_{\cal B} \sqrt{B(x)} w_n(x)\sqrt{\gamma}d^{n-2}y=
{1 \over 2} \int_{\cal B} T dN~~
\end{equation}
equals half of the mass of the source.

As in (\ref{7}), the quantity
$dN=d\sigma/G$ is identified with the number of degrees of freedom per element
of the screen with the area $d\sigma=\sqrt{\gamma}d^{n-2}y$.
The `equipartition' form of (\ref{35}), see \cite{Padmanabhan:2009kr,Verlinde:2010hp}, also
supports the choice of the thermodynamical relation in the form (\ref{32}).

It is instructive to discuss also the case when the massive source is a Schwarzschild black hole and the minimal surface goes near its horizon.
There is a point where the distance between the surface and the horizon is minimal. At this point the normal vector to $\cal B$ is directed along the acceleration
of a static observer. In the limit when the surface is shifted closer to the horizon the local temperature (\ref{34}) tends to the Hawking temperature $T_H=\kappa / 2\pi$ where $\kappa=\partial_rB(r_H)/2$ is the surface gravity and the derivative is taken along the radial coordinate. This is another evidence
for the validity of equations (\ref{32}) and (\ref{34}).

We see, therefore, that at least some variational formulas of minimal surfaces allow a thermodynamical
analogy. It is a subject of further studies to find a deeper meaning behind this analogy. We also see
that minimal surfaces considered as holographic screens provide a strong support to the hypothesis that gravity has an entropic origin \cite{Verlinde:2010hp}.

Because the physical interpretation of the entropy associated to minimal surfaces is known one can pose
problems for future research. We mention two of them.
First, it is important to understand a possible role of quantum effects which may uncover the meaning of the
local temperature (\ref{34}), like quantum evaporation of black holes explains the meaning of
the Hawing temperature.

Second, it is important to study generalization of equation (\ref{0})
when the low energy limit of the gravity theory differers from the Einstein form.
The simplest example of this situation are higher curvature corrections which may be present in the gravity action.
The problem is that  higher curvature terms change both formula (\ref{1})
and equations which determine the shape of  $\cal B$.
If this generalization allows a thermodynamical analogy the ideas discussed here are robust.

%\bibliographystyle{fullcream} % for the author's convenience
%\bibliography{df}

\end{document}